# Nanoanalytical TEM studies and micromagnetic modelling of Nd-Fe-B magnets

Gregor A. Zickler[1], Peter Toson[1], Ahmad Asali[1] and Josef Fidler[1]

[1]*Institute of Solid State Physics, TU Wien, Vienna, Austria*
*gregor.zickler@tuwien.ac.at*

**Abstract**

We have analysed the influence of the microstructural features, such as intergranular grain boundary (GB) phases and misalignment of the hard magnetic grains, on the optimization of magnetization reversal processes in order to improve the coercive field of Nd-Fe-B magnets.

The microstructural model of the grains and intergranular phases, which is used for theoretical simulations, has been derived from a detailed nanoanalytical TEM/STEM study of a Dy/Tb free magnet and a high coercive (Nd,Tb)-Fe-B magnet. Special attention is laid on the EELS analysis of GB with a thickness ranging from 2 - 30 nm. This analysis identified the majority of the GB phases to have about 50 -70 at.% of iron and only a few GBs, which are connecting two nearby grain boundary junctions (GBj), possess a similar chemical composition as the adjacent GBj with a low iron content (< 10 at. %) and a high rare earth and oxygen content.

Finite element micromagnetic simulations have been carried out in order to study the influence of internal demagnetizing fields determined by the microstructure on the magnetization switching behaviour. Special emphasis was put on the influence of the GB and their magnetic properties, due to their substantial influence on the nucleation of reverse magnetic domains and the pinning of domain walls. The strongest reduction of the coercive field is caused by GB with soft ferromagnetic properties. Shielding the Nd-Fe-B grains from the nucleation sites at the GBj with Dy or Tb shells, leads to an increase of the coercivity from 2.5 to 3.6 T and 2.5 to 4.3 T, respectively.

# 1 Introduction

The search for candidates of suitable magnetic materials and their expected behaviour as reduction of the dysprosium/terbium content is of great economic and scientific interest [1]–[4]. Micromagnetic simulations have been widely used for the interpretation and visualization of experimental results. Additionally, the correct prediction of hard magnetic hysteresis properties and reversal mechanisms of new materials should be the aim [5]–[7]. Nanoanalytical and high resolution transition electron microscopic investigations of various Nd-Fe-B sintered magnets with different rare earth (RE) content and coercive field, were the input for the creation of a numerical finite element model used in micromagnetic simulations. Special emphasis was laid on the chemical composition of the grain boundary phases. To understand the role of local variations of the saturation polarization and magnetocrystalline anisotropy, as they appear at grain boundaries, and to understand the impact of the addition of heavy rare earth elements, like dysprosium and terbium, on the resulting coercive field are the aim of these numerical simulations.

## 2 Experimental and micromagnetic simulation procedure

The chemical and morphological investigations from Nd-Fe-B permanent magnets were carried out on an analytical field emission transmission electron microscope (TEM) (FEI Tecnai F20) at 200 kV, which is equipped with a silicon drift energy dispersive X-ray detector from EDAX, a Gatan Tridem GIF electron energy loss spectrometer (EELS) and a high angle annular dark field detector (HAADF). TEM specimens were prepared using the lift-out technique in a focused ion beam (FIB) (FEI Quanta 200 3D DualBeam-FIB).

The software package FEMME, which is a hybrid finite element / boundary element method code, was used to compute the micromagnetic simulations [8]. The program solves the Landau-Lifshitz-Gilbert equation [9] on a magnetic volume, which is discretized with finite elements. The uniaxial magnetocrystalline anisotropy constant $K_1$, the saturation polarization $J_s$ and the exchange stiffness A are input parameters for the simulation. The easy axis of a volume can be rotated with a polar angle $\theta$ and an azimuthal angle $\varphi$. Soft ferromagnetic materials (grain boundaries) can be simulated setting $K_1$ to zero. For paramagnetic materials $K_1$ and A can be set to zero and for nonmagnetic materials, which represent weak dia- or paramagnets, $J_s$ can be set to a small but non-zero value.

The most intensive part of the simulation, computationally, is the calculation of the demagnetizing field $H_{mag}$. An exact analytical solution exists only for special geometries, like rotational ellipsoids. The demagnetizing field acts as additional nucleation mechanism, especially on the surface of the simulated model, where high values occur. It also takes the influence of the shape anisotropy on the coercivity in account. Simulations, where the demagnetizing field is switched off, are labeled $H_{mag}$=off. Only the exchange and magnetocrystalline anisotropy are the basis for the resulting coercivity in these simulations. The comparison of simulation with $H_{mag}$=off and $H_{mag}$=on, creates the possibility of quantifying the influence of the demagnetizing field.

## 3 Results

### 3.1 TEM investigation

Nanoanalytical and selected area electron diffraction (SAED) TEM/STEM techniques were used to determine the chemical composition and morphology of the phases occurring in sintered Nd-Fe-B permanent magnets. Sample (a) was produced with He jet milling in order to obtain the smallest possible grain size of 1.4 μm and it contains only neodymium and praseodymium and no heavy rare earth (HRE) elements ($\mu_0 H_{cJ} \approx 2.1$ T). Sample (b) has 2.4 wt% of Tb in its starting composition and the standard milling gas nitrogen was used resulting in an average grain size of 3.4 μm ($\mu_0 H_{cJ} \approx 3.0$ T). Figure 1 shows HAADF images of the FIB lamella of sample (a) (a-1) and of sample (b) (a-2).

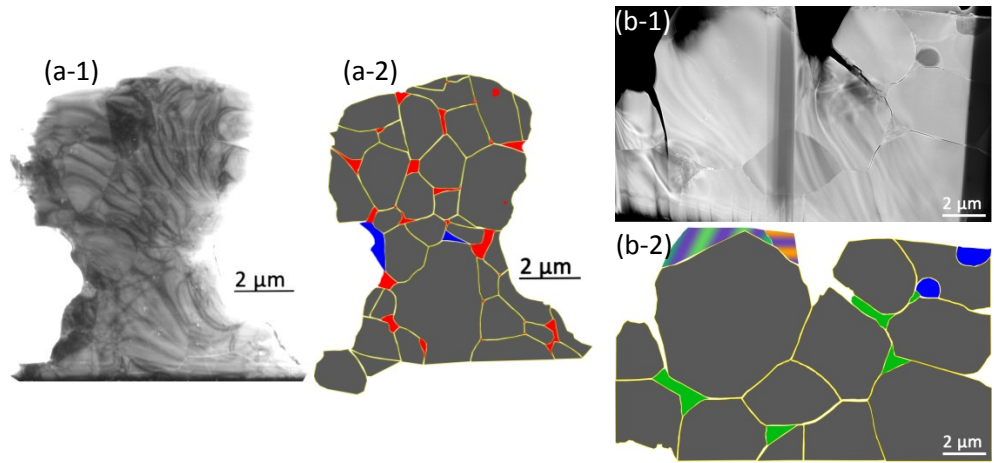

**Figure 1:** STEM image of a FIB lamella of a HRE-free sintered Nd-Fe-B magnet (a-1) and of a sintered (Nd,Tb)-Fe-B magnet (b-1). (a-2) and (b-2) are morphological maps, where specific colors were assigned to the occurring phases.

Several RE-rich phases were found beside the main 2-14-1 phase, which is colored gray in Figure 1. The c-$Nd_2O_3$ phase, which has a cubic crystal structure (a = 1.108 nm, $Ia\bar{3}$ (no.206)) [10], is mostly found in grain boundary junctions (GBj) and is colored red in Figure 1 (a-2). This phase was identified with a Fast Fourier Transformation (FFT) of a high resolution TEM (HR-TEM) image (Figure 2 ②&③). The fcc-NdO phase (a = 0.4994 nm, $cF$8 (no.225)) [11] occurs in grain like structures in sample (a) and was identified with SAED images (Figure 2 ④) acquired in the blue phase in Figure 1 (a-2). These phases were also reported by Fidler [12], Fidler et al. [13], Tang et al. [14], Lemarchand et al. [15], Wang and Li [16] and Hrkac et al. [17]. The RE-rich GBj phase of sample (b), which is colored in green, is a mixture of the $Nd_2O_3$ and NdCu phase. A bright field TEM image showing a GBj and three neighboring $Nd_2Fe_{14}B$ grains, which are separated by GB phases, is displayed in Figure 2 ①. A GB which is perpendicular to the c-axis of the adjacent grain is indicated as "x"-GB and one that is parallel is a "y"-GB. Figure 2 ② shows a HR-TEM image of the edge of the GBj, were the FFT for the identification of the GBj-phase was taken from.

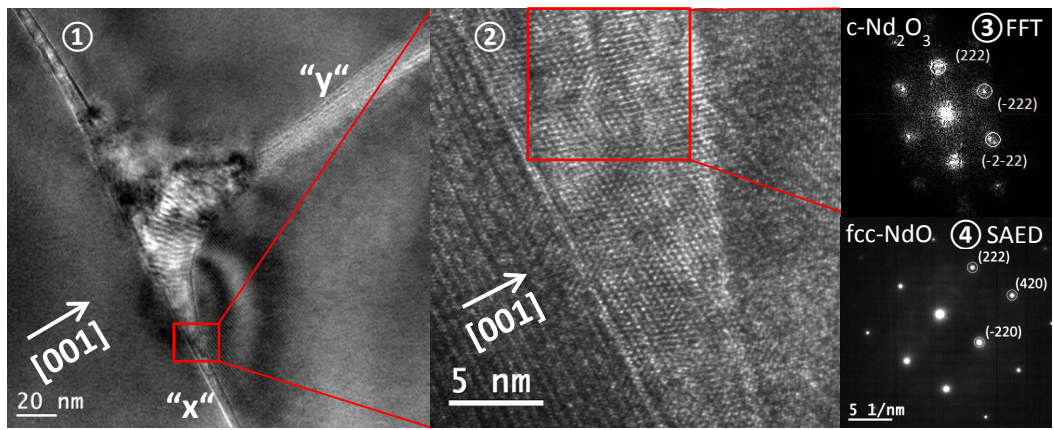

**Figure 2:** Bright field image of a GBj surrounded by three $Nd_2Fe_{14}B$ grains ①, HR-TEM image of the edge of the GBj ②, FFT accrediting the c-$Nd_2O_3$ phase to the GBj ③, SAED of an fcc-NdO RE-rich phase ④.

Detailed EELS experiments were conducted, in order to study the chemical composition of GB in both samples. For reliable quantitative EELS analysis, $Nd_2Fe_{14}B$, $Pr_2Fe_{14}B$ and $Tb_2Fe_{14}B$ standards were investigated. The high quality of the chemical analysis of these samples can be seen in the good correspondence of the measured composition of the gray phase $RE_{11.6}Fe_{82.3}B_{6.1}$ with respect to the nominal composition $RE_{11.8}Fe_{82.3}B_{5.9}$ of the 2:14:1 phase. A HR-TEM image showing an "x"-GB, where the [001] and [110] lattice plains of the left grain are well visible, is displayed in Figure 3①. The HAADF image on the bottom of ① attributes the thickness of this GB to be 2.2 nm. The average thickness of the GB in sample (a) was 3.1 nm and 18.7 nm in sample (b). Figure 3 ② and ③ show the change in the chemical composition along an EELS line-scan from a $RE_2Fe_{14}B$ grain over a GB and into another grain. Two different kinds of GBs were found in these samples. The majority of the GB have a high iron content of 50 – 70 at. % (Figure 3 ②), which is in good agreement with recent three dimensional atom probe (3D-ATP) experiments of sintered Nd-Fe-B magnets [18], [19]. GB connecting two nearby GBj, possess a similar chemical composition as the adjacent GBj with a low iron content (< 10 at. %) and a high RE and oxygen content (Figure 3 ③).

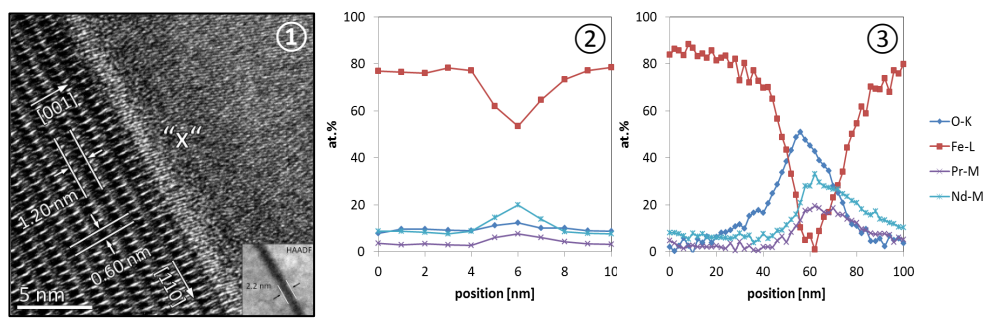

**Figure 3:** HR-TEM image of an "x"-GB ①, EELS line-scan from a grain across a GB and into another grain ② and ③.

## 3.2 Micromagnetic simulations

In order to determine the influence of GB phases on the nucleation of reverse magnetic domains in the presence of an external magnetic field in sintered Nd-Fe-B magnets, micromagnetic simulations with various models have been carried out (Figure 4). A Model consisting of $Nd_2Fe_{14}B$ grains separated by a 4 nm GB, which is orientated perpendicular to the easy axis of magnetization (c-axis) of the adjacent grains ("x"), was used to examine the role of the GB as a nucleation site of reverse magnetic domains. Another similar Model, where the GB is orientated parallel to the c-axis of the adjacent grains, was used to determine the pinning behavior of the GB. The 8-g model ① (Figure 4) consists of 8 100x100x100 nm $Nd_2Fe_{14}B$ grains separated by a 4 nm GB phase and is a combination of the two models mentioned before. The misorientation angle $\theta_o$ of the c-axis of the hard magnetic grains with respect to the external field was varied, to obtain realistic conditions for the simulation. The influence of a core-shell (cs) structure, like in Dy-F treated Nd-Fe-B magnets, on $H_{cJ}$ is simulated in the 8-g-cs model ② (Figure 4), where an 8 nm thick $HRE_2Fe_{14}B$ (HRE=Dy, Tb) shell was put between the GB and the $Nd_2Fe_{14}B$ grains.

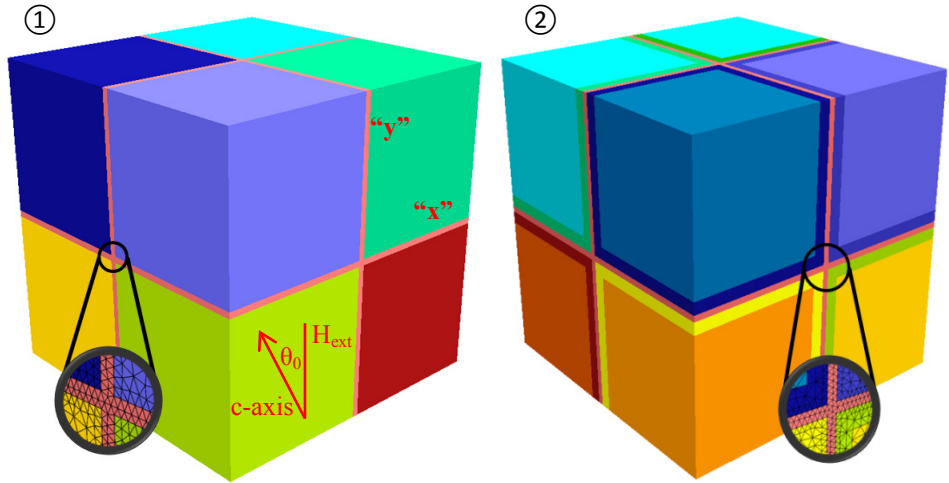

**Figure 4:** Finite element models used in micromagnetic simulations consisting of 8 $Nd_2Fe_{14}B$ grains separated by a 4 nm GB ①, the grains are separated from the GB by an 8 nm thick $HRE_2Fe_{14}B$ ②.

For a ferromagnetic (fm) GB the nucleation site of the reverse magnetic domains (blue) is located at region with the highest density of the GB material, which is at the GBj in the center of the model where the three GB plains intersect (Figure 5, ①). For small values of $\theta_o$ the expansion of this reverse domain is isotropic along the GB and into the hard magnetic grains. In case of paramagnetic (pm) GB the nucleation site of the reverse magnetic domains occurs at the GBj, only if the demagnetising field $h_{mag}$ is turned off. The nucleation site moves to the GBj on the surface, which is perpendicular to the external field, if $h_{mag}$=on. The nucleation sites of the reverse magnetic domains in the model with nonmagnetic (nm) GB are located at the outer corners of the model, where the field strength of the demagnetizing field reaches its maximum value (Figure 5, ②). The pinning behaviour of the "y"-GB is visible at the top left grain in ②. The strong influence of the magnetic properties of the GB and the demagnetizing field on the coercive field $H_{cJ}$ can be seen in the hysteresis loop, where the continuous lines correspond to $h_{mag}$=on and the dashed lines to $h_{mag}$=off (Figure 5, ③). The orientation of the spins in the ferro magnetic GB is coupled to the surrounding hard magnetic grains due to the high exchange stiffness A (see Table 1). In case of the paramagnetic GB, A is much smaller and in combination with the high $J_S$, this GB is not coupled with the grains and switches already with a still positive external field, due to a reduction of the demagnetizing field. The nonmagnetic GB, having a low A and $J_S$, is therefore hardly influenced by the external magnetic field and switches simultaneously with the grains. The simulation with nm GB and $h_{mag}$=off results in an $H_{cJ}$, which reaches 98 % of the anisotropy field of $Nd_2Fe_{14}B$ (7.65 T). The nm GB acts as a magnetic shield, prohibiting any magnetic exchange between the grains. If $h_{mag}$=on, the grains are exchange coupled through the demagnetizing field, reducing $H_{cJ}$ by almost 2 T. In the model with fm GB $h_{mag}$ has the smallest influence, because the grains are already strongly coupled due to the high exchange stiffness A of the fm GB.

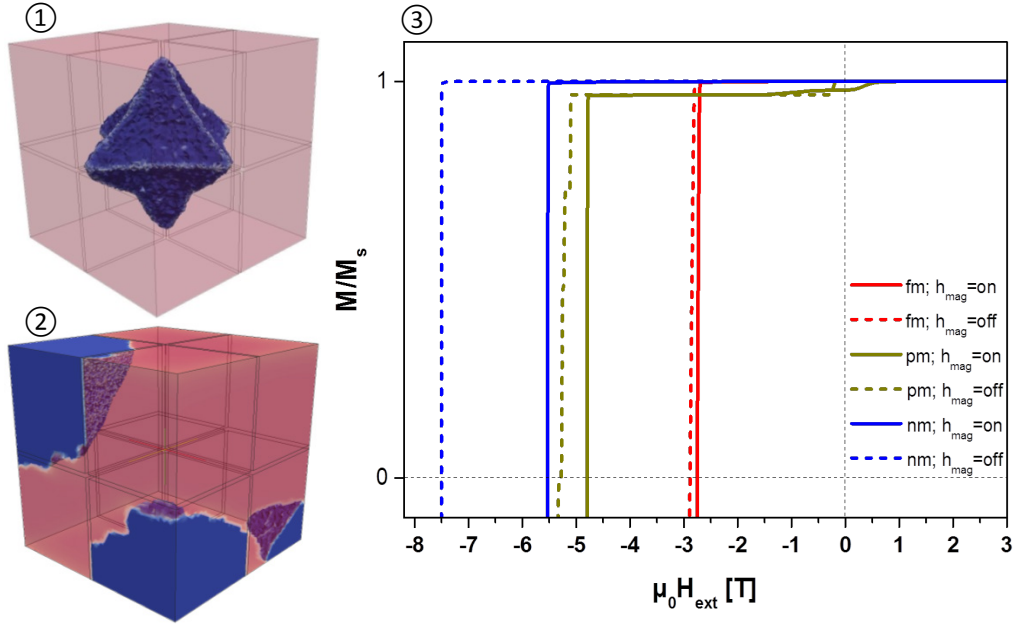

**Figure 5:** For ferromagnetic (fm) GB nucleation of a reverse magnetic domains occurs at the center of the 8-g model ①, the reverse magnetic domains nucleate at the edges of the model in case of nonmagnetic (nm) GB and $h_{mag}$=on ②. Simulated hysteresis loops of different GB materials and $h_{mag}$=off (dashed) & on is shown in ③.

The material parameters of the GB phases are shown in Table 1. $K_1$ was set to zero in all the GB materials. Recent electron holographic TEM and spin-polarized SEM experiments attribute the GB to be ferromagnetic with a $J_s$ of 0.5 – 1.0 T [20], [21]. Therefore the $J_s$ of the fm GB was set to 0.75 T. The high Fe content of the GB, which was obtained by the EELS and 3D-ATP experiments [18], [19], indicates the GB to be soft-ferromagnetic. Therefore the exchange stiffness A was set to approximately a tenth of the value bulk iron (22 pA/m) [22] in fm GB and one hundredth of the value of $Nd_2Fe_{14}B$ (A=7.7 pA/m) [23] in nm and pm GB. Comparable resulting $H_{cJ}$ were obtained by molecular dynamic- and finite element calculations [24]. Here the grains were surrounded by a thin defect layer with low anisotropy and separated by a nonmagnetic GB.

The strong influence of local fluctuations of $K_1$ and $J_s$ on $H_{cJ}$, like in the model with HRE-shell with an fm GB, is shown in Table 1. The models with nm and pm GBs have a stronger dependence on the misorientation angle, especially for small angles. The influence of the misorientation $\theta_o$ of the grains on $H_{cJ}$, in models with nm and pm GB, is higher than the addition of the HRE layer. Therefore the values of the coercive field are higher without a HRE shell and $\theta_0 = 0°$, than with a HRE shell and $\theta_0 = 10°$.

| structure | non-mag. ($\theta_0$) $J_s$ = 0.001 T $A$ = 0.077 pA/m | para-mag. ($\theta_0$) $J_s$ = 0.75 T $A$ = 0.077 pA/m | soft-ferro-mag. ($\theta_0$) $J_s$ = 0.75 T $A$ = 2.5 pA/m |
|---|---|---|---|
| 8-g | 5.54 (0°) | 4.79 (0°) | 2.75 (0°) \| 2.46 (10°) |
| 8-g-cs-Dy | 4.65 (10°) | 4.65 (10°) | 3.58 (10°) |
| 8-g-cs-Tb | 4.66 (10°) | 4.66 (10°) | 4.33 (0°) \| 4.34 (10°) |

**Table 1:** Overview of the resulting coercive field $H_{cJ}$ in T of various models and different grain boundary properties ($h_{mag}$=on). Material parameter: $Nd_2Fe_{14}B$: $K_1$ = 4.9 MJ/m³, $J_s$ = 1.61 T, A = 7.7 pJ/m **[23]**, $Dy_2Fe_{14}B$: $K_1$ = 4.5 MJ/m³, $J_s$ = 0.67 T, A = 7.7 pJ/m **[25]**, $Tb_2Fe_{14}B$: $K_1$ = 6.13 MJ/m³, $J_s$ = 0.7 T, A = 7.7 pJ/m [2].

# 4  Conclusion

TEM experiments have attributed the size of the hard magnetic grains of the investigated samples to be 1.4 – 3-4 µm. The mesh size of the finite element model used in micromagnetic simulations has to be in the order of the exchange length of the simulated material, which is 1-2 nm for $Nd_2Fe_{14}B$. This limits the maximum possible grain size to 100 – 200 nm, resulting in several million finite elements. Therefore a micromagnetic finite element model (200 nm)³ coupled with a macroscopic finite element model (1-3 µm)³, based on Maxwell equations, will be used to obtain realistic results for sintered Nd-Fe-B magnets [26]. The main advantage of this approach is the reduction of the demagnetizing field on the surface of the micromagnetic model.

The presence of a soft-ferromagnetic GB phase in sintered Nd-Fe-B magnets is supported by the following facts:

- A high iron content of the GB, which was confirmed by EELS (Figure 3) and 3D-APT experiments [18], [19].
- Recent electron holographic TEM and spin-polarized SEM experiments accredit the GB to be soft-ferromagnetic with a $J_s$ of 0.5 – 1.0 T and significant exchange [20], [21].
- The simulation of the 8-g model with a ferromagnetic GB and $\theta_o = 10°$ results in an $H_{cJ} = 2.46$ T, which is not far from the measured $H_{cJ} = 2.1$ T of sample (a). Simulation with para- and nonmagnetic GB overestimate $H_{cJ}$ by 2.69 T and 3.44 T, respectively (Table 1)

The increase of $H_{cJ}$ by 0.9 T from sample (a) to sample (b) can be attributed to the following reasons:

- The presence of HRE elements between the GB phase and the hard magnetic grains, acts as a protective shield against the nucleation of reverse magnetic domains and acts as a pinning layer to prevent domain wall movement through the material.
- GB with a high RE and oxygen content were found, which should have para- or nonmagnetic properties and therefore have a positive effect on $H_{cJ}$ (Figure 5, ③).

# Acknowledgements

The funding from the European Community's Seventh Framework Programme (FP7-NMP) under Grant agreement no. 309729 (ROMEO) is acknowledged. TEM investigation was carried out using facilities at USTEM, TU Wien. The authors are grateful to Drs. M. Katter and K. Uestuener from Vacuumschmelze GmbH and Co KG for providing the magnet samples.

# References

[1] O. Gutfleisch, M. A. Willard, E. Brück, C. H. Chen, S. G. Sankar, and J. P. Liu, "Magnetic Materials and Devices for the 21st Century: Stronger, Lighter, and More Energy Efficient," *Adv. Mater.*, vol. 23, no. 7, pp. 821–842, Feb. 2011.

[2] J. F. Herbst, "R 2 Fe 14 B materials: Intrinsic properties and technological aspects," *Rev. Mod. Phys.*, vol. 63, no. 4, pp. 819–898, Oct. 1991.

[3] J. J. Croat, J. F. Herbst, R. W. Lee, and F. E. Pinkerton, "Pr-Fe and Nd-Fe-based materials: A new class of high-performance permanent magnets (invited)," *J. Appl. Phys.*, vol. 55, no. 6, p. 2078, 1984.

[4] H. Yamamoto, Y. Matsuura, S. Fujimura, and M. Sagawa, "Magnetocrystalline anisotropy of R2Fe14B tetragonal compounds," *Appl. Phys. Lett.*, vol. 45, no. 10, p. 1141, 1984.

[5] R. Fischer and H. Kronmüller, "The role of the exchange interaction in nanocrystalline isotropic Nd2Fe14B-magnets," *J. Magn. Magn. Mater.*, vol. 191, no. 1–2, pp. 225–233, Jan. 1999.
[6] D. Suess, M. Kirschner, T. Schrefl, J. Fidler, R. Stamps, and J.-V. Kim, "Exchange bias of polycrystalline antiferromagnets with perfectly compensated interfaces," *Phys. Rev. B*, vol. 67, no. 5, Feb. 2003.
[7] H. Sepehri-Amin, T. Ohkubo, M. Gruber, T. Schrefl, and K. Hono, "Micromagnetic simulations on the grain size dependence of coercivity in anisotropic Nd–Fe–B sintered magnets," *Scr. Mater.*, vol. 89, pp. 29–32, Oct. 2014.
[8] T. Schrefl, D. Suess, W. Scholz, H. Forster, V. Tsiantos, and J. Fidler, "Finite Element Micromagnetics," in *Computational Electromagnetics*, vol. 28, P. Monk, C. Carstensen, S. Funken, W. Hackbusch, and R. H. W. Hoppe, Eds. Berlin, Heidelberg: Springer Berlin Heidelberg, 2003, pp. 165–181.
[9] T. L. Gilbert, "Classics in Magnetics A Phenomenological Theory of Damping in Ferromagnetic Materials," *IEEE Trans. Magn.*, vol. 40, no. 6, pp. 3443–3449, Nov. 2004.
[10] P. Villars and L. D. Calvert, *Pearson's handbook of crystallographic data for intermetallic phases*, 2nd ed., vol. Vol. 4. Materials Park, OH: ASM International, 1991.
[11] J. M. Leger, N. Yacoubi, and J. Loriers, "Synthesis of rare earth monoxides," *J. Solid State Chem.*, vol. 36, no. 3, pp. 261–270, Mar. 1981.
[12] J. Fidler, "Analytical microscope studies of sintered Nd-Fe-B magnets," *IEEE Trans. Magn.*, vol. 21, no. 5, pp. 1955–1957, Sep. 1985.
[13] J. Fidler and K. G. Knoch, "Electron microscopy of Nd-Fe-B based magnets," *J. Magn. Magn. Mater.*, vol. 80, no. 1, pp. 48–56, Aug. 1989.
[14] W. Tang, S. Zhou, R. Wang, and C. D. Graham, "An investigation of the Nd-rich phases in the Nd-Fe-B system," *J. Appl. Phys.*, vol. 64, no. 10, p. 5516, 1988.
[15] D. Lemarchand, P. Vigier, and B. Labulle, "On the oxygen stabilized Nd-rich phase in the Nd-Fe-B (-O) permanent magnet system," *IEEE Trans. Magn.*, vol. 26, no. 5, pp. 2649–2651, Sep. 1990.
[16] S. C. Wang and Y. Li, "In situ TEM study of Nd-rich phase in NdFeB magnet," *J. Magn. Magn. Mater.*, vol. 285, no. 1–2, pp. 177–182, Jan. 2005.
[17] G. Hrkac, T. G. Woodcock, K. T. Butler, L. Saharan, M. T. Bryan, T. Schrefl, and O. Gutfleisch, "Impact of different Nd-rich crystal-phases on the coercivity of Nd–Fe–B grain ensembles," *Scr. Mater.*, vol. 70, pp. 35–38, Jan. 2014.
[18] H. Sepehri-Amin, Y. Une, T. Ohkubo, K. Hono, and M. Sagawa, "Microstructure of fine-grained Nd–Fe–B sintered magnets with high coercivity," *Scr. Mater.*, vol. 65, no. 5, pp. 396–399, Sep. 2011.
[19] H. Sepehri-Amin, T. Ohkubo, T. Shima, and K. Hono, "Grain boundary and interface chemistry of an Nd–Fe–B-based sintered magnet," *Acta Mater.*, vol. 60, no. 3, pp. 819–830, Feb. 2012.
[20] Y. Murakami, T. Tanigaki, T. T. Sasaki, Y. Takeno, H. S. Park, T. Matsuda, T. Ohkubo, K. Hono, and D. Shindo, "Magnetism of ultrathin intergranular boundary regions in Nd–Fe–B permanent magnets," *Acta Mater.*, vol. 71, pp. 370–379, Jun. 2014.
[21] T. Kohashi, K. Motai, T. Nishiuchi, and S. Hirosawa, "Magnetism in grain-boundary phase of a NdFeB sintered magnet studied by spin-polarized scanning electron microscopy," *Appl. Phys. Lett.*, vol. 104, no. 23, p. 232408, Jun. 2014.
[22] J. M. D. Coey, *Magnetism and magnetic materials*, Repr. Cambridge: Cambridge Univ. Press, 2013.
[23] R. Skomski and J. M. D. Coey, *Permanent magnetism*. Bristol, UK ; Philadelphia, PA: Institute of Physics Pub, 1999.
[24] G. Hrkac, K. Butler, T. G. Woodcock, L. Saharan, T. Schrefl, and O. Gutfleisch, "Modeling of Nd-Oxide Grain Boundary Phases in Nd-Fe-B Sintered Magnets," *JOM*, vol. 66, no. 7, pp. 1138–1143, Jul. 2014.
[25] R. Skomski, *Simple models of magnetism*. Oxford ; New York: Oxford University Press, 2008.
[26] F. Bruckner, D. Suess, M. Feischl, T. Führer, P. Goldenits, M. Page, D. Praetorius, and M. Ruggeri, "Multiscale modeling in micromagnetics: Existence of solutions and numerical integration," *Math. Models Methods Appl. Sci.*, vol. 24, no. 13, pp. 2627–2662, Dec. 2014.